\documentclass[prd,aps,psfig,floats]{revtex4}

\usepackage[pctex32]{graphics}
\begin{document}

\title{{\bf Density perturbations in warm inflation and COBE normalization}}




\author{H. P. de Oliveira}
\email{oliveira@dft.if.uerj.br}
\affiliation{{NASA/Fermilab Astrophysics Center \\
Fermi National Accelerator Laboratory, Batavia, Illinois, 60510-500.}\\
and \\
{\it Universidade do Estado do Rio de Janeiro }\\
{\it Instituto de F\'{\i}sica - Departamento de F\'{\i}sica Te\'orica,}\\
{\it CEP 20550-013 Rio de Janeiro, RJ, Brazil}}



\begin{abstract} 
Starting from a gauge invariant treatment of perturbations an analytical
expression for the spectrum of long wavelength density perturbations in warm
inflation is derived. The adiabatic and entropy modes are exhibited
explicitly. As an application of the analytical results, we determined the
observational constraint for the dissipation term compatible with
COBE observation of the cosmic microwave radiation anisotropy for some specific
models. In view of the results the feasibility of warm inflation is discussed. 
\end{abstract}

\maketitle


\section{Introduction}

Inflation\cite{kolbtur} is the better description of the early stages of the
Universe, where  besides providing a satisfactory solution for the main
problems of the standard Cosmology, also predicts a mechanism to generate the
inhomogeneities for the structure formation. The advent of more precise
observational data given by COBE\cite{cobe}, MAXIMA\cite{maxima}, etc has
so far confirming the basic predictions of the inflation. Nonetheless, since
the first model of inflation was proposed, we have witnessed a myriad of
alternative models, as for instance, new inflation, chaotic inflation, extended
inflation, hybrid inflation, warm inflation, etc. A direct way of deciding
which model must be discarded is confronting their theoretical predictions with
the appropriate observational data\cite{turner,kolbdodel}. In this context, I
shall discuss the warm
inflation\cite{berefang,olivrud,gleiser,bellini,leefang1,taylor,linde,olivjor}
(WI) whose important feature is that thermal fluctuations during inflation play
a dominant role in producing the initial spectrum of perturbations. In WI
radiation is continuously generated through a dissipation mechanism resulting
from the interaction of the inflaton with other fields. Therefore, it is
expected that at the end of inflation enough radiation for a smooth transition
to the next phase, and the reheating phase is no longer necessary.   


Warm inflation can be understood as a two-field inflation model. In this case,
entropy perturbations arise from the variation of the effective equation of state
relating the total pressure with the total energy density. As a consequence,
contrary to the single-field inflation models, the
curvature perturbation\cite{bardeen,mukhanov} $\zeta$\ has a non-trivial
evolution after the perturbation crosses outside the horizon, and the way in
which it evolves in super-horizon scales depends on the details of the model
itself. We expect that COBE normalization provides observational limits on the
dissipation term $\Gamma$ once, as we have already shown\cite{olivjor}, the
dissipation term plays a crucial role in producing the entropy
perturbations. Recently, de Oliveira and Joras\cite{olivjor} investigated
cosmological perturbations in WI together with the first determination of the
observational  limits on the dissipation term\footnote{Berera and
Fang\cite{berefang} have established an expression for the range of $\Gamma$ in
the weak dissipative regime just by considering $T_r>H$, $T_r$ being the
temperature of the thermal bath, without any treatment on the evolution
of cosmological perturbations.}. Moreover, in this determination it was assumed
a very small dissipation in the sense of allowing, as a good approximation, the
conservation of $\zeta$. Under such circumstances, we found $\Gamma \sim
10^{-16}\,m_{Pl}$ for two distinct models, being compatible with other
estimations\cite{olivrud,leefang1} which the only criterion was the requirement
that at the end of inflation there is enough radiation for the next phase. On
the other hand, it will very important to determine the observational limits on
$\Gamma$ with no assumption whether the dissipation is small or not when
compared with the Hubble parameter. Then, this observational test will
constitute a definitive solution of the theoretical dispute\cite{gleiser,linde}
concerning the feasibility of WI.

This paper is divided as follows. In Section II the basic equations that
describe WI, and the equation for the perturbations are presented. Taking into
account the slow-roll conditions, an approximate solution of these
equations is derived where the contribution of adiabatic and entropy modes are
shown explicitly. In Section III, we present a procedure to normalize the
amplitude of density perturbations in WI using the COBE data. A specific
application of COBE normalization is performed in Section IV. Finally, in
Section V is devoted to discuss the main results and conclusions.

\section{Adiabatic and entropy  modes in warm inflation}

The basic equations that describe the homogeneous background dynamics of
interacting radiation and scalar fields in flat Friedmann-Robertson-Walker
spacetimes are\cite{berefang}  

\begin{eqnarray}
& & H^2=\frac{8\pi}{3m^2_{Pl}}\, 
\left( \rho_r +\frac{1}{2}{\dot\phi}^2+V(\phi)\right)  \\
& & \ddot \phi +(3H+\Gamma)\dot \phi + V^\prime(\phi)=0, \\
& & \dot \rho_r + 4 H \rho_r =  \Gamma {\dot\phi}^2. 
\end{eqnarray}

\noindent  Here $H=\frac{\dot{a}}{a}$ is the Hubble parameter, $m_{Pl}$ is 
the Planck mass, dot and prime denote derivative with respect to the 
cosmological time $t$ and the
scalar field, respectively; $\rho_r$ is the energy density of the radiation
field. The dissipative term $\Gamma$ is responsible for the decay of
 the scalar field into radiation during the slow-roll phase. In general,
$\Gamma$ can be assumed as a function of $\phi$ of the type $\Gamma = \Gamma_m
\phi^m$, reflecting the types of interactions between the scalar field $\phi$
and other fields\cite{berefang,olivrud}. In addition, $m$ must be even and
$\Gamma_m > 0$ to guarantee the positiveness of $\Gamma$ as demanded by the
Second Law of Thermodynamics. We remark that there is a theoretical 
controversy about the introduction of the dissipative term in
Eq. (2) during the slow-roll\cite{gleiser,linde}, and therefore on the
feasibility of WI. Our approach here will be to assume it as part of the WI
scenario and, from a consistent treatment of cosmological perturbations,  allow
that observation dictates the result of this issue.    

Warm inflation is characterized by the accelerated growth of the scale factor
driven by the potential term $V(\phi)$ that dominates over other energy terms in
Eq. (1), or 

\begin{equation}
H^2 \simeq \frac{8\pi}{3m^2_{Pl}}\,V(\phi).\label{eq4}
\end{equation} 

\noindent Also, the $\ddot{\phi}$ term can be neglected in Eq. (2), yielding 

\begin{equation}
\dot{\phi} \simeq -\frac{V^{\prime}(\phi)}{3 H + \Gamma}, \label{eq5}
\end{equation}

\noindent and the radiation density is kept approximately constant given by

\begin{equation}
\rho_{r} \simeq \frac{\Gamma \dot{\phi}^2}{4 H}. \label{eq6}
\end{equation}

\noindent Several authors have studied the dynamics of WI in which the basic 
issue was to present WI as a viable description of the early
universe. In this context, it was shown that even for a tiny dissipation, in the
sense of the ratio $\alpha \equiv \frac{\Gamma}{3 H} \ll 1$, 
enough radiation is found at the end of inflation for a smooth
transition to the radiation era. In particular, for $\alpha$ of order $10^{-9}$
is in agreement with observational data\cite{olivjor}.

We consider inhomogeneous perturbations of the FRW background described by
the metric in the longitudinal gauge\cite{bardeen,mukhanov} 

\begin{equation}
d s^2 = (1+2 \Phi)\,d t^2 - a^2(t) (1-2 \Psi) \delta_{i j}\,d x^i\,d x^j,
\label{eq7}
\end{equation}

\noindent where $a(t)$ is the scale factor, $\Phi=\Phi(t,\bf{x})$ and
$\Psi=\Psi(t,\bf{x})$ are the metric perturbations. The spatial dependence of all
perturbed quantities are of the form of plane waves $e^{i\bf{k.x}}$, $k$
being the wave number, so that the perturbed field equations regarding now only
their temporal parts (we omit the subscript $k$) are

\begin{eqnarray}
& & \dot{\Phi} + H \Phi = \frac{4 \pi}{m_{pl}^2} \left(-\frac{4}{3 k} \rho_r a v
 + \dot{\phi} \delta \phi \right) \label{eq8}  \\
& & (\delta \phi \ddot{)} + (3 H + \Gamma)\,(\delta \phi \dot{)} + \left(
\frac{k^2}{a^2} + V^{\prime\prime} + \dot{\phi}
\Gamma^{\prime}\right)\,\delta \phi = 4  
\dot{\phi} \dot{\Phi} + (\dot{\phi} \Gamma - 2 V^{\prime})\, \Phi \label{eq9} \\
& & (\delta \rho_r \dot{)} + 4 H \delta \rho_r + \frac{4}{3} k a \rho_r v = 
4 \rho_r \dot{\Phi} + \dot{\phi} \Gamma^{\prime}
\delta \phi + \Gamma \dot{\phi}\,(2 (\delta \phi\dot{)} - 3 \dot{\phi}
\Phi) \label{eq10} \\
& & \dot{v} + \frac{\Gamma \dot{\phi}^2}{\rho_r}\,v + \frac{k}{a}\,
\left(\Phi + \frac{\delta \rho_r}{4 \rho_r} + \frac{3 \Gamma \dot{\phi}}
{4 \rho_r}\,\delta
\phi \right) = 0 \label{eq11}
\end{eqnarray}

\noindent where $\delta \rho = \delta \rho_r + \dot{\phi}\,(\delta \phi \dot{)}
- \dot{\phi}^2\,\Phi + V^{\prime}\,\delta \phi$ and $\delta p = \frac{1}{3}
\delta \rho_r + \dot{\phi}\,(\delta \phi \dot{)} - 
\dot{\phi}^2\,\Phi - V^{\prime}\,\delta \phi$ are the perturbations of the
total energy density and pressure, respectively; $v$ originates from the
decomposition of the velocity field as $\delta U_i=-\frac{i a
k_i}{k}\,v\,e^{i \bf{k.x}}$ (see Bardeen\cite{bardeen}). Also, due the fact
that the perturbation of the total energy-momentum tensor does not give rise to
anisotropic stress ($\delta T^i_j \propto \delta^i_j$), $\Psi=\Phi$.  Warm
inflation can be considered as a hybrid-like inflationary model since two
basics fields, the inflaton and the radiation fields,  are interacting during
the slow-roll phase. Therefore, isocurvature or entropy perturbations are
generated, besides the adiabatic ones. We have shown\cite{olivjor} that the
entropy  perturbations are directly related to the dissipation term. In what
follows, we will obtain an approximate solution for the long wavelength
perturbations valid during the slow-roll phase, in which the adiabatic and the
entropy modes are exhibited explicitly.    

During the slow-roll phase, it can be assumed that the perturbed quantities do
not vary strongly (see the appendix). This means, that in Eq. (\ref{eq8})  $H
\Phi \gg \dot{\Phi}$; in Eq. (\ref{eq9}), $(\delta \phi \ddot{)} \ll (\Gamma +
3 H) (\delta \phi \dot{)}$, and  so on. Following these approximations, 
together with the slow-roll conditions provided by Eqs. (\ref{eq4}),
(\ref{eq5}) and (\ref{eq6}), $\delta \rho_r$ and $v$ are expressed as 

\begin{equation}
\frac{\delta \rho_r}{\rho_r} \simeq \frac{\Gamma^{\prime}}{\Gamma}\,\delta \phi
- 3 \Phi, \label{eq12}
\end{equation}

\begin{equation}
v \simeq -\frac{k}{4 a H}\,
\left(\Phi + \frac{\delta \rho_r}{4 \rho_r} + \frac{3 \Gamma \dot{\phi}}
{4 \rho_r}\,\delta \phi \right). \label{eq13}
\end{equation}

\noindent Using these equations, the metric perturbation (Eq. (\ref{eq8})) can be
written as

\begin{equation}
\Phi \simeq \frac{4 \pi \dot{\phi}}{m_{pl}^2 H} \left(1+\frac{\Gamma}{4 H} +
\frac{\Gamma^{\prime} \dot{\phi}}{48 H^2} \right)\,\delta \phi. \label{eq14}
\end{equation}

\noindent Note that in the case of null dissipation, we recover the standard
relation between the metric and scalar field perturbations, $\Phi \simeq
\frac{4 \pi \dot{\phi}}{m_{pl}^2 H}\,\delta \phi$, that describes the adiabatic
mode\cite{linde1,plastaro}. 

The equation of motion for $\delta \phi$ reads now as

\begin{equation}
(\Gamma + 3 H)\,(\delta \phi \dot{)} + (V^{\prime\prime} + \dot{\phi}
\Gamma^{\prime})\,\delta \phi \simeq (\dot{\phi} \Gamma -2
V^{\prime})\,\Phi, \label{eq15}
\end{equation}

\noindent where it will be useful to introduce an auxiliary function $\chi$ by

\begin{equation}
\chi = \frac{\delta \phi}{V^{\prime}}\,exp\left(\int
\frac{\Gamma^{\prime}}{\Gamma + 3 H} d \phi\right), \label{eq16}
\end{equation}

\noindent that generalizes the procedure due to Starobinski and
Polarski\cite{plastaro} in the realm of double inflationary models. Substituting
Eq. (\ref{eq16}) into (\ref{eq15}), we obtain the following equation for $\chi$

\begin{equation}
\frac{\chi^{\prime}}{\chi} + \frac{9}{8}\,\frac{(\Gamma + 2 H)}{(\Gamma + 3
H)^2}\,\left(\Gamma + 4 H - \frac{\Gamma^{\prime} V^{\prime}}{12 H (\Gamma + 3
H)}\right)\,\frac{V^{\prime}}{V} \simeq 0 \label{eq17} 
\end{equation}

\noindent We separate two situations: $\Gamma = \Gamma_0 = const$, and the
variable dissipation term $\Gamma = \Gamma_m \phi^m$, $m = 2,4,..$. Considering
the first case, Eq. (\ref{eq17}) can be integrated exactly, yielding 

\begin{equation}
\delta \phi \simeq \frac{C_1 \dot{\phi}}{H}\,\left(1 + \frac{\Gamma_0}{3
H}\right)^{5/4}\,exp\left[\frac{\Gamma_0}{4 (\Gamma_0 +3 H)}\right]. \label{eq18}
\end{equation}

\noindent where $C_1$ is a constant of integration. It will be instructive to
expand this expression in power series of $\Gamma_0$, 

\begin{equation}
\delta \phi \simeq \frac{C_1 \dot{\phi}}{H}\,\left(1 + \frac{\Gamma_0}{2 H}
+\frac{\Gamma_0^2}{36 H^2} + ...\right). \label{eq19}
\end{equation}

\noindent Notice that the zeroth order term is the same found in
single field inflationary models\cite{linde1,plastaro}, which is related the
adiabatic mode. The remaining terms describe the effect of dissipation to the
fluctuation of the scalar field, or in another words, the entropy mode. The
metric perturbation is obtained straightforwardly from Eqs. (\ref{eq14}) and
(\ref{eq18}), 

\begin{equation}
\Phi \simeq - \frac{C_1 \dot{H}}{H^2}\,\left(1 + \frac{\Gamma_0}{4
H}\right)\,\left(1 + \frac{\Gamma_0}{3
H}\right)^{1/4}\,exp\left[\frac{\Gamma_0}{4 (\Gamma_0 + 3 H)}\right]. \label{eq20}
\end{equation}
  
\noindent Again, expanding it power series of $\Gamma_0$, the role of the
dissipation in producing entropy mode becomes evident

\begin{equation}
\Phi \simeq - \frac{C_1 \dot{H}}{H^2}\,\left(1 + \frac{5 \Gamma_0}{12 H} +
\frac{\Gamma_0^2}{72 H^2} + ...\right).  \label{eq21}
\end{equation}

\noindent For the second case, the integration of (\ref{eq17}) can be performed
only after specifying the potential $V(\phi)$.  

An important relation between the density of matter fluctuations, $\frac{\delta
\rho}{\rho}$, and the metric perturbation can be derived in WI after taking into
account that, during the slow-roll phase, $\frac{\delta \rho}{\rho} \simeq
\frac{V^{\prime}}{V} \delta \phi$, and Eq. (\ref{eq14}),     

\begin{equation}
\frac{\delta \rho}{\rho} \simeq -\frac{8}{3}\,\frac{(\Gamma + 3
H)}{\left(\Gamma + 4 H + \frac{\Gamma^{\prime} \dot{\phi}}{12 H}\right)}\,\Phi.
\label{eq22}  
\end{equation}

\noindent In the absence of the dissipation, we recover the usual relation
$\frac{\delta \rho}{\rho} \simeq -2 \Phi$ valid for single field inflationary
models. On the other limit, for instance, high constant dissipation, $\Gamma_0
\gg H$, it follows that  $\frac{\delta \rho}{\rho} \simeq - \frac{8}{3}
\Phi$. The above relation together with the integration of Eq. (\ref{eq17})
will be of fundamental importance for the COBE normalization as we are going to
show next.

\section{COBE normalization}

In super cooled inflation the overall dynamics of perturbations can be reduced
to a single conservation law for the gauge invariant curvature perturbation on
comoving hypersurfaces, $\zeta$, defined by\cite{bardeen,mukhanov}

\begin{equation}
\zeta \equiv \frac{2 \rho}{3 H (\rho + p)}\,(H \Phi + \dot{\Phi}) +
\Phi. \label{eq23} 
\end{equation}

\noindent During inflation, we have $H \Phi \gg \dot{\Phi}$ such that, together
with Eqs. (\ref{eq18}), (\ref{eq20}) and (\ref{eq22}) and $\Gamma = 0$, it
follows that the curvature perturbation can be approximated as  $\zeta \simeq
\frac{\delta \rho}{\rho+p}$\cite{linde1,turner1}. In order to illustrate the 
importance of this quantity, consider a perturbation corresponding the size of
a galaxy, $\lambda_{gal}$. Accordingly, it can be shown that this perturbation
leaves the horizon about  50 e-folds before the end of inflation. Then,
$\zeta_{50} \simeq \left(\frac{\delta \rho}{\rho+p}\right)_{50}$, with  the
subscript 50 indicating that the curvature perturbation is evaluated at 50
e-folds before the end of inflation, is frozen until the perturbation re-enters
the Hubble horizon in the radiation or matter dominated eras. When this
happens, it can be established that $\left(\frac{\delta
\rho}{\rho+p}\right)_{50} \propto \left(\frac{\delta
\rho}{\rho}\right)_{horizon,\lambda}$ is the amplitude of density perturbations
on the scale $\lambda_{gal}$ when the perturbations crosses back inside the
horizon during the post-inflation epoch, where the constant of proportionality
depends whether $\rho+p=n \rho$, $n=1,4/3$ for matter, radiation era,
respectively. For instance, for single-field inflation models, it can
be show that\cite{turner,kolbdodel} $\left(\frac{\delta
\rho}{\rho}\right)_{horizon,\lambda} \simeq \left(\frac{V^{3/2}}{m^3_{Pl}
V^{\prime}}\right)_{50}$, where quantum fluctuations are the  source of the
perturbations of the scalar field. The Sachs-Wolfe effect\cite{sachswolfe}
establishes, roughly speaking, that the metric perturbation, which in turn is
related to the density perturbations, determines the temperature anisotropy of
the CBR (cosmic microwave background), or 

\begin{equation}
\frac{\delta T}{T} \simeq
\left(\frac{\delta \rho}{\rho}\right)_{horizon,\lambda}. \label{eq24}  
\end{equation}

\noindent The COBE satellite performed reliable measurements of the anisotropies
of temperature, and therefore, became a powerful tool of testing inflation
models by normalizing the amplitude of perturbations, as well as determining
the spectral index of the spectrum of perturbations.

Under the influence of entropy perturbations, the conservation of $\zeta$ in
super-horizon scales  is no longer valid, and in the specific case of WI, the
rate of variation of $\zeta$ is depends directly on the dissipation
term\cite{olivjor}. An important observational test for WI is the determination
of the magnitude of the dissipation term compatible with COBE data, implying
that it can be decided if strong dissipative regime characterized by $\Gamma
\gg H$, weak dissipative regime $\Gamma \ll H$, or even dissipation of the
order of the Hubble parameter is the most viable model for WI. Indeed, this
task was partially accomplished by de Oliveira and Joras\cite{olivjor} by
assuming {\it ab initio} weak dissipation such that the effect of the entropy
perturbation could be neglected. Under this assumption it was possible to
determine a lower bound of the dissipation term taking into
account that thermal fluctuations are mainly responsible for the production of
the inhomogeneities of the inflaton, and the COBE data, where it was found that
$\Gamma \approx 10^{-16}\,m_{Pl}$. 

Here, we derive a convenient expression for $\zeta_{end}$ necessary for COBE
normalization, in which it is assumed that the dissipation is constant and
vanishes after inflation. This 
last condition is physically reasonable, since dissipation results from the
interaction between the inflaton and other fields, and at the end of inflation
the inflaton field has transferred the enough energy to radiation to start a
new phase without the necessity of reheating.  The next step is to obtain a
suitable expression for the curvature perturbation $\zeta$ in WI. For this, we
may consider that during the slow-roll phase, Eq. (\ref{eq23}) is reduced to  

\begin{equation}
\zeta \simeq \frac{2 \rho}{3 (\rho+p)}\,\Phi \label{eq26}
\end{equation}

\noindent due to the quasi-static evolution of the perturbations in this
phase. This equation together with Eq. (\ref{eq22}) yield,

\begin{equation}
\zeta \simeq - \frac{(\Gamma_0 + 4 H)}{4 (\Gamma_0 + 3 H)}\,\frac{\delta
\rho}{\rho + p}. \label{eq27}
\end{equation}

\noindent In the limit $\Gamma_0 = 0$, as well as for $\Gamma_0 \gg H$, $\zeta
\propto \frac{\delta \rho}{\rho + p}$, suggesting that the effect of the dissipation
is to change this proportionality. Now, we are in conditions trace out a
procedure to apply the COBE normalization.

Consider again a perturbation in WI corresponding to a scale of astrophysical
interest that corresponds to the size of a galaxy. This perturbation will
cross outside the Hubble radius at approximately 50 e-folds before the end of
inflation\cite{olivjor}. As previously discussed, the presence of entropy
perturbations are responsible for the variation of $\zeta$ in super-horizon
scales, and such a variation takes place until the end of inflation, since
$\Gamma_0=0$ after inflation. The curvature perturbation
$\zeta_{end}$ associated to this perturbation is kept constant until the moment
when the perturbation re-enters the Hubble horizon during the post-inflation
epoch to become the necessary fluctuations for structure formation. The
amplitude of the resulting spectrum is normalized by COBE data. As
one can see, we need to know $\zeta_{end}$ instead of $\zeta_{50}$ due to the
variation of curvature perturbation. To determine the former, we calculate the
ratio

\begin{equation}
K \equiv \frac{\zeta_{end}}{\zeta_{50}} \simeq
\left(\frac{\rho}{\rho+p}\right)_{end}\,\left(\frac{\rho}{\rho+p}\right)^{-1}_{50}\,
\frac{\Phi_{end}}{\Phi_{50}}, \label{eq28}
\end{equation}

\noindent where we have taken into account Eq. (\ref{eq26}). The ratio between the
metric perturbation follows from the approximate solution (\ref{eq20}), for which
it can be shown that for $\Gamma_0 = 0$, $K = 1$, as expected. Then, the
expression for $\zeta_{end}$ as given by 

\begin{equation}
\zeta_{end} = K\,\zeta_{50} \simeq - K \left[\frac{(\Gamma_0 + 4 H)}{4
(\Gamma_0 + 3 H)}\,\frac{\delta \rho}{(\rho+p)}\right]_{50}, \label{eq29}
\end{equation}

\noindent that must be normalized by COBE data. Notice that $\delta \rho \simeq
V^{\prime}(\phi)\,\delta \phi$, with the fluctuations of the scalar field being
due to thermal interactions with the radiation field\cite{berefang}. In
the next Section we apply the COBE normalization to determine the dissipation
term. 


\section{Worked examples: high dissipation and COBE data} 

Let us consider the case of quadratic potential $V(\phi)=\frac{1}{2} m^2
\phi^2$. It will be useful to introduce adimensional quantities such as 

\begin{equation}
\gamma_0 = \frac{\Gamma_0}{m},\;\; x = \frac{\sqrt{4 \pi}}{m_{Pl}} \phi,\;\;
\alpha = \frac{\Gamma_0}{3 H} \simeq \frac{\gamma_0}{\sqrt{3} x}, \label{eq30} 
\end{equation}

\noindent where in the last expression it is assumed that $H^2 \simeq \frac{8
\pi}{3 m_{Pl}^2} V(\phi)$. The regime of high dissipation is  characterized by
$\alpha \gg 1$. The end of WI is achieved when $\epsilon_{wi} \simeq 
\frac{m_{Pl}^2}{4 \pi \alpha} \left(\frac{H^{\prime}}{H}\right)^2 = 1$
for $x = x_e$, where $\epsilon_{wi}$ is the generalized slow-roll parameter
for WI\cite{olivjor}. The beginning of WI at $x=x_*$ can be determined from the
condition  of 60 e-folds necessary for a successful inflation. Then, both
conditions render  

\begin{equation}
x_{e} \simeq \frac{\sqrt{3}}{\gamma_0},\:\:x_* \simeq \frac{61 \sqrt{3}}{\gamma_0}.
\label{eq31}
\end{equation}

\noindent To guarantee that the regime of high dissipation holds since the
beginning of WI, it is necessary that $\gamma_0^2 \gg 183$. Also, if $x_{50}$
corresponds the scalar field evaluated at 50 e-folds before the end of
inflation, it can be shown that  

\begin{equation}
x_{50} \simeq \frac{51 \sqrt{3}}{\gamma_0}.
\label{eq34}
\end{equation}


According to the discussion of the last Section, we need to know $K$, which can
be determined from the approximate solution for $\Phi$ given by Eq.
(\ref{eq20}) together with the slow-roll conditions (\ref{eq4}), (\ref{eq5}),
(\ref{eq6}). The derived expression is found to be

\begin{equation}
K \simeq
\left(\frac{\alpha_{e}}{\alpha_{50}}\right)^{5/4} \approx 1.36 \times 10^2,
\end{equation}

\noindent where the values of $\alpha$ at the end of inflation and at 50 e-folds 
before the end of inflation are given by Eqs. (\ref{eq31}) and (\ref{eq34}).
This value of $K$ is the maximum reached by considering very high dissipation.
The next step is to consider Eq. (28) assuming $\delta \rho \simeq
V^{\prime}(\phi) \delta \phi$, with the fluctuations of the scalar field being
generated by thermal interactions with radiation, instead of quantum
fluctuations. Then, following Berera and Taylor\cite{taylor}, it is established
that

\begin{equation}
(\delta \phi)^2 \simeq \frac{(\Gamma_0 H)^{1/2} T_r}{2 \pi^2}, \label{eq35}
\end{equation}

\noindent  where $T_r$ is the temperature of the thermal bath. These
fluctuations are evaluated at 50 e-folds before the end of 
inflation. After performing this calculation, together with Eqs. (\ref{eq30}),
(\ref{eq31}), (\ref{eq34}) and the condition $\alpha \gg 1$, we arrived to a simple
expression for $\zeta_{end}$ 

\begin{equation}
|\zeta_{end}| \simeq 93.45  \left(\frac{m}{m_{PL}}\right)^{3/4} \gamma_0^{3/4}.
\end{equation}

\noindent COBE normalization tells us that  $|\zeta_{end}| \simeq 7.33
\times 10^{-6}$, implying 
 
\begin{equation}
\Gamma_0 \simeq 3.36 \times 10^{-10} m_{PL}, \label{eq37}
\end{equation}

\noindent which, in view of our approximation, this result can be interpreted
as the upper bound of the dissipation term. In order to assure the regime of high
dissipation during the whole WI, it is necessary that $\gamma_0 \gg 14$. This
restriction together with the obtained upper bound of $\Gamma_0$ implies that

\begin{equation}
m \ll 2.40 \times 10^{-11} m_{PL}.
\label{eq38}
\end{equation} 

\noindent This constraint is a very small bound for the mass of the
inflaton, even if compared with the value obtained by COBE normalization in
super-cooled inflation\cite{turner}, which is $m \sim 10^{-6}\,m_{PL}$.



We have considered another potential $V(\phi)=\lambda^4 \phi^4$ and variable
dissipation term $\Gamma=\Gamma_2 \phi^2$, with $\Gamma_2$ being a constant of
dimension $m_{PL}^{-1}$. In this case, high dissipation is characterized by

\begin{equation}
\alpha \simeq \frac{m_{PL}
\Gamma_2}{\sqrt{24 \pi} \lambda^2} \gg 1.
\end{equation}

\noindent The scalar field evaluated at the end, the beginning  and at 50
e-folds before the end of WI, correspond, respectively, to

\begin{equation}
x_e \simeq \frac{2}{\sqrt{\alpha}},\; x_* \simeq \sqrt{61} x_e,\;x_{50} \simeq
\sqrt{51} x_e. \end{equation}

\noindent For the sake of completeness, the end of WI occurs for $\epsilon
\simeq \frac{m_{PL}^2}{\pi \alpha} \frac{1}{\phi^2}$, and the remaining values
are determined from $N \simeq \frac{\pi \alpha}{m_{PL}^2} (\phi_*^2-\phi_N)$,
by setting $N=50,\,60$. We need to know $\delta x$ in order to determine the
metric perturbation from Eq. (\ref{eq14}), and consequently $K$ (see Eq.
(\ref{eq26}) which is valid for any $\Gamma$). After integrating Eq.
(\ref{eq17}) imposing $\alpha \gg 1$, the perturbation of the scalar field is
given by

\begin{equation}
\delta x \simeq 4 \lambda^4 \chi_0 x^{-7/2} exp\left(-\frac{3}{4\alpha^2
x^2}\right), \label{eq40}
\end{equation}

\noindent where $\chi_0$ is the constant of integration. Taking into account
this result and the corresponding expression for $\Phi$, we obtain

\begin{equation}
K \simeq \left(\frac{3 \alpha x_e^2-1}{3 \alpha x_{50}^2-1}\right)
\frac{x_{50}}{x_e} \frac{(\delta x)_e}{(\delta x)_{50}} \approx 1.24 \times
10^2.
\end{equation}

\noindent Notice that the maximum value of growth of the curvature
perturbation is approximately the same as found in the previous case. The final
step for the COBE normalization is to consider Eq. (\ref{eq35}) inserted into

$\zeta_{end} = K \zeta_{50} \simeq -K \left[\frac{\left(\Gamma + 4 H +
\frac{\Gamma^{\prime} \dot{\phi}}{12 H}\right)}{4 (\Gamma + 3 H)}
\frac{V^{\prime} \delta \phi}{(\rho+p)}\right]_{50}$ together with the value of
$K$ given by Eq. (\ref{eq40}). Hence, after a direct calculation

\begin{equation}
|\zeta_{end}| \simeq 3.21 \times 10^3 \lambda^{3/2}.
\end{equation}

\noindent Contrary to the previous case, the dissipation term does not appear
explicitly. COBE normalization imposes that

\begin{equation}
\lambda^4 \simeq 10^{-24}.
\end{equation}


\section{Discussion}

In this paper, we have integrated the equations for the metric and scalar 
field perturbation in WI. The contributions of the adiabatic and entropy 
modes were exhibited explicitly showing the central role played by the 
dissipation in producing the entropy mode. Another very important result was 
the generalization of the relation between the total density of matter
perturbations and metric perturbations given by Eq. (\ref{eq22}). Using the
analytical expressions, a general procedure to apply COBE  normalization was
proposed, since in WI a conservation equation for the curvature perturbation,
$\zeta$, is no longer valid. 

As an application of the COBE normalization, we have considered  two distinct
examples, namely, quadratic potential $V(\phi)=\frac{1}{2} m^2 \phi^2$ with
constant dissipation, and quartic potential $V(\phi)=\lambda^4 \phi^4$ with
variable dissipation term. The common aspect shared by these two examples is
the assumption of high dissipative regime guaranteed by
$\frac{\Gamma}{3 H} \gg 1$. In the first case, COBE normalization determines
directly the upper bound of the dissipation term as $\Gamma_0 \simeq 3.36 \times
10^{-10}\,m_{PL}$, with the mass of the inflaton given by $m \ll 2.40 \times
10^{-11} m_{PL}$. Thus, considering the previous result concerning lower bound
of the dissipative term\cite{olivjor}, the following constraint is obtained

\begin{equation}
\Gamma_{min} \leq \Gamma_0 \leq \Gamma_{max}
\end{equation} 

\noindent where $\Gamma_{min}/m_{PL} \approx 10^{-16}$ and
$\Gamma_{max}/m_{PL} \approx 10^{-10}$. On the other hand, for the
case of quartic potential, COBE normalization imposes that $\lambda^4 \simeq
10^{-24}$, which represents a stronger fine tuning if compared with $10^{-14}$
found in supercooled inflation. In spite of representing a drawback for very
high dissipative regime in WI in this specific example, it is necessary further
analysis considering other situations in order to give a final word about the
feasibility of the very high dissipative regime of WI as far as COBE
normalization is concerned. In this way, a possible direction of investigation
is to treat other cases characterized by distinct potentials and by not
imposing {\it a priori} any kind of high/low dissipative regime to verify the
consequences of COBE normalization.

\section{Acknowledgments}

I acknowledge CNPq for financial support. I am  grateful to Rocky
Kolb and J. Frieman for useful discussions. Finally, I am also grateful to the
anonymous referee whose observations improved the final version of the
manuscript.


\appendix
\section{Appendix}

It is well established that the slow-roll phase $\dot{H} \ll H^2$. This
condition has a direct consequence on the evolution of the perturbed quantities
as we are going to illustrate integrating Eq. (\ref{eq8}). The general solution
of Eq. (\ref{eq8}) can be written as

\begin{equation}
\Phi = \frac{\Phi_0}{a} + \frac{1}{a}\,\int \Delta a d t
\end{equation}

\noindent where $\Phi_0$ is the initial value of the metric perturbation and
$\Delta = \frac{4 \pi}{m_{pl}^2} \left(-\frac{4}{3 k} \rho_r a v  + \dot{\phi}
\delta \phi \right)$  for convenience. The non-decreasing mode is given by the
integral, that can be expanded in power series\cite{mukhanov,linde1} like

\begin{equation}
\Phi = \frac{\Delta}{H} - \left(\frac{\Delta}{H}\right)^{.}\,H^{-1} +
\left[\left(\frac{\Delta}{H}\right)^{.}\,H^{-1}\right]^{.}\,H^{-1} + ...
\end{equation}

\noindent Then, it is acceptable to approximate the above expression
considering only the first term of the series, since during the slow-roll,
$\frac{\dot{H}}{H^2} \ll 1$ and $\frac{\dot{\Delta}}{H^2} \ll 1$. As we can
see, this approximation is equivalent to assume $H \Phi \ll
\dot{\Phi}$. Applying this procedure to the other perturbed equations, we
arrive to the Eqs. (\ref{eq12}), (\ref{eq13}), (\ref{eq14}) and (\ref{eq15}).

\end{document}